# The conceptual and mathematical foundations of the MC-QTAIM

*Shant Shahbazian*

*Department of Physics, Shahid Beheshti University, Tehran, Iran*
*E-mail: sh_shahbazian@sbu.ac.ir*

**Abstract**
The concept of "atoms in molecules" (AIM) is one of the cornerstones of the structural theory of chemistry. However, in contrast to the free atoms, a comprehensive quantum mechanical theory of AIM, treating them as quantum particles or quantum subsystems, has never been proposed. Currently, the most satisfactory deduction of this concept is based on the "partitioning" methodologies that are trying to recover AIM from the ab initio wavefunctions. One of these methodologies is the quantum theory of AIM (QTAIM), which retrieves AIM by an exhaustive partitioning of the one-electron density into atomic basins in real space. The molecular properties are then partitioned into the basin and inter-basin contributions as the incarnation of the AIM properties and their interaction modes, respectively. The inputs of the QTAIM partitioning scheme are the electronic wavefunctions computed from the electronic Schrödinger equation, which is a "single-component" equation treating electrons as quantum particles and the nuclei as clamped point charges. A recently extended form of the QTAIM, called the multi-component QTAIM (MC-QTAIM), removes this restriction and enables AIM partitioning to be applied to the MC many-body quantum systems. This is done using MC wavefunctions as inputs that are derived from the MC Schrödinger equation in which there are two or more types of quantum particles. This opens the possibility for the AIM partitioning of molecular systems where certain nuclei, e.g. because of their non-adiabatic coupling to electrons, must be treated as quantum particles instead of clamped point charges. The same formalism allows the partitioning of exotic molecular systems in which there are other elementary particles like muons or positrons, in addition to electrons and nuclei. The application of the MC-QTAIM partitioning to such systems reveals that the positively charged muon may shape its atomic basin, i.e. an example of "exotic AIM", while a positron may act as an agent of bonding, i.e. an example of "exotic bonds".

**Key points/objectives box:**
The concept of "atoms in molecules" is extended to multi-component quantum systems.
The extended formalism is applied to "atoms in molecules" partitioning of the muonic and positronic molecular systems.
The partitioning demonstrates that the positively charged muon forms its own atomic basin in the muonic molecules.
The partitioning demonstrates that the positron may act as a bonding agent and forms a novel positronic bond with an unprecedented mechanism.

# I. Introduction: Why is there a problem with "atoms in molecules"?

In his succinct but marvelous paper on the meaning of emergence and its implications for the condensed matter physics, Blundell, a condensed matter physicist, borrows Conway's Game of life as a toy model of an artificial universe with simple but well-defined laws,[1] to demonstrate his thesis on the reality of the emergent entities.[2] The heart of this thesis is the premise that upon the interaction of elementary agents of a many-agent system at a lower/microscopic level, novel agents at a higher level may emerge, with their own identities.[2] As he emphasizes rightfully, these higher-level coarse-grained agents can be much more complex and diverse than the elementary ones, but they are as real as the latter. While it is probably easier to compute the state and time evolution of a many-agent system at the lower level, the higher one is much more effective for "storytelling". A story is a casual narrative that is comprehensible to the human mind and when told, triggers the "Aha erlebnis", but defies a formalization and is not compressible into a mathematical expression like a dynamical law. Briefly, we may compute the properties of a system/phenomenon at the lower level but think and try to understand the same system/phenomenon at the higher level. Although the target of Blundell is digging into the nature of modern theoretical condensed matter physics, his description of the methodological intricacies of that field and the conclusions he has drawn may seem quite familiar to a theoretical chemist as well. Indeed, a proper solution to the so-called "atoms in molecules" (AIM) problem in theoretical chemistry, as will be discussed, requires an understanding of the emergence and the appreciation of the need for storytelling in chemistry.

The modern chemical language is the result of a long and complex evolution that has its roots in the mid-19th century when the structural theory of molecules was first proposed, and rapidly became the central dogma of chemistry.[3] In the original structural theory, each molecule has a chemical structure that consists of the constituent atoms and the network of the bonds linking the atoms in the molecule. While in this original version chemical structures were merely "topological" entities,[4] i.e. equal to modern mathematical graphs where nodes are the AIM and links are the bonds,[5–7] subsequent developments transformed chemical structures into "geometrical" arrangements of atoms and bonds in 3D space.[8] At the dawn of the 20th century, since the discovery of electrons and nuclei, atoms lost their status as elementary agents, and the theory transformed into the electronic structural theory to cope with these discoveries.[9–12] The Lewis-Langmuir atomic theory with its cubically arranged static electrons within atoms and the shared electron pairs as the incarnation of bonds between atoms was the precursor of this new age,[13–16] while the subsequent developments had to await the discovery of modern quantum mechanics (QM).[17] This new age finally established the fundamental level of modern electronic structure theory, let us call it the *lower level* in which Schrödinger's equation governs molecules containing the properties and interaction modes of electrons and nuclei as the elementary agents of chemistry.[18] There is no mention of AIM at the lower level anymore, and if indeed one is only interested to compute various observables of a molecule, the concept of AIM is in fact redundant and unnecessary. As stressed masterfully by Laughlin and Pines,[19] Schrödinger's equation is the "theory of everything" for almost all chemistry and condensed matter physics.

However, the original concept of AIM based on the idea that the free atoms preserve their identities within molecules not only did not fade away but also flourished in modern chemistry. The chemical bond is probably the most central concept in modern chemistry,[20–23] but, how is it possible to conceive a bond without conceiving the two or more atoms in a molecule that the bond links them? The basic classification scheme in organic chemistry is according to the concept of functional group,[24,25] i.e. a few atoms in a molecule that collectively yield a novel and distinct emergent entity. Meanwhile more recently clusters and superatoms are gaining the same role in inorganic and materials chemistry.[26–31] Thus, not only AIM and their properties and interactions are of concern as an *intermediate level* of organization in modern structural chemistry, but even *higher levels* of the structural chemistry emerge upon certain arrangements of AIM. As if matter arranges itself in ways that novel collective identities emerge at various scales of length and complexity and chemistry is understandable only by storytelling using these higher-level entities.

All these impart a dichotomy in modern chemical language. On the one hand, modern electronic structure theory of molecules yields its conceptual tools to comprehend chemical systems and phenomena, e.g. the atomic and molecular orbitals,[32,33] selection rules for chemical reactions,[34,35] and various electronic effects.[36] On the other hand, chemists do not restrict themselves only to these lower-level concepts, and emergent concepts at the intermediate level are ubiquitous in chemical language, e.g. the oxidation states of atoms,[37] the atomic charges and bond energies,[38] and the interatomic interaction potentials.[39] Many questions arise for a theoretical chemist that tries to put all these in a coherent framework: What are AIM? Particles, subsystems, or…? How the AIM and their properties are related to the properties of electrons and nuclei, and are they deducible from or reducible to the lower level? Is it possible to describe a molecule quantum mechanically and predict its properties just by employing the properties and interaction modes of AIM, instead of those of electrons and nuclei? Is there a universal dynamical quantum equation, similar to Schrödinger's equation, governing AIM? These and many other similar questions are not new and have been posed in different frameworks by thoughtful theoretical chemists.[40–43] In fact, all these are part of a larger and increasingly tense question, with a rich technical and philosophical literature,[44–66] on how various levels in physics, chemistry, biology, and even the human sciences are interrelated? Let us first have a look at the possible quantum mechanical frameworks that one may employ to define AIM.

## II. How to introduce a theoretical framework for AIM?

The idea that at the most fundamental level the universe is made of ingredients that have fixed, context-independent, properties, and interaction modes is hardwired into the programs that are trying to find the so-called "theory of everything".[67] This is also the case for the "theory of everything" of chemistry and condensed matter physics as discussed previously,[19] since, electrons and nuclei have fixed properties, e.g. their masses and charges, and their interaction modes are universal, e.g. the Coulomb law. However, in theoretical condensed matter physics, a new type of particles emerges, called quasiparticles, where their properties like mass or charge are imposed by the environment in which they are formed and sometimes they are derivable through a mathematical procedure called the *renormalization*.[68–70] What about AIM? If one tries to imagine AIM as some sort of quantum particles, they seem more like quasiparticles rather than the fundamental particles. In fact, the prime facet of the classical models like molecular mechanics, which are trying to start from AIM and their interactions to predict molecular potential energy surfaces, is the concept of "atom type".[71–73] This asserts that an atom in various chemical environments has different properties and interaction modes much like the situation of quasiparticles. On the other hand, the theory of interatomic interactions also reveals that there is no universal law for the atom-atom interaction potentials and the explicit form of these potentials heavily depends not only on the pair of atoms but also on their environment, i.e. their surrounding atoms.[39,74] Accordingly, if one insists to contemplate AIM as quantum particles, they are "shape shifter" particles or to borrow the terminology used by Blundell,[2] they are *emergent particles*. Nevertheless, in contrast to the case of quasiparticles, currently, there is no systematic mathematical procedure like the mentioned renormalization approach to derive AIM and their types from the lower-level theory.

An alternative view is to assume that AIM are somehow quantum subsystems. If a molecule is a quantum system, then AIM may be conceived as clusters of electrons around each nucleus while the number of electrons of each cluster can be fixed or allowed to vary.[75] To proceed in this direction, the system must be decomposed into quantum subsystems, i.e. AIM, in a way that each subsystem has its own Hilbert space, and the Hilbert space of the system will be the tensor product of the Hilbert spaces of the subsystems.[76] This is only feasible if there is a subsystem Hamiltonian, and its corresponding eigenfunctions are used as the basis set to construct the subsystem Hilbert space.[77,78] However, electrons are *indistinguishable* quantum particles and it is hard to conceive how one may group them into subsets in a molecule and then try to write a subsystem Hamiltonian for each subset.[76] Also, if one conceives that electrons are fluctuating between AIM, which is a quite reasonable assumption as will be demonstrated, then the whole idea of attributing a fixed number of electrons to each subset seems completely misguided.[75] Therefore, as far as it is not possible to describe each atom in a molecule with its own *distinguishable* internal "degrees of freedom", whatever they are, it is hard to imagine how the standard theoretical apparatus, developed for the open quantum subsystems,[77,78] applies to them.

At the current state of affairs, the only remaining option is trying to somehow "partition" a molecule into AIM by employing *ad hoc* methodologies that are "reasonable" algorithms to divide the state of the system, i.e. wavefunction, or certain state-derived functions, e.g. the reduced density matrices or the one-electron densities, into AIM "contributions".[79–82] As mentioned previously, the usual partitioning in QM is the partitioning of the Hilbert space of the system into the product of subspaces, and this tacitly implies that there are certain quantum subsystems within the system. Thus, without such subsystems, any other proposed partitioning, while maybe mathematically sound and chemically straightforward to interpret, is not *uniquely* derivable from the principles of QM without certain auxiliary *ad hoc* principles.[83–85]

The outcome of a partitioning algorithm is the contribution of each AIM to the molecular properties, e.g. their share from the total energy and the number of electrons, as well as their share from various types of intramolecular interactions, e.g. the bond energies. If applied to a certain molecule, because of their *ad hoc* nature, the contributions derived from different partitioning algorithms are not the same quantitatively, or sometimes even qualitatively, and this makes cross-examinations between these algorithms quite non-trivial.[86] Currently there are two general trends for developing the partitioning schemes, which may be called *bottom-up* and *top-down* methodologies. In both methods, the starting point is the solution of the molecular Schrödinger equation, i.e. the molecular wavefunction. In the bottom-up methods, the mathematical representation of the wavefunction is represented in a manner that separate clusters of electrons around each nucleus, conceived as AIM, have distinct representatives in the wavefunction.[87–93] This is not a trivial task since electrons are indistinguishable particles, though, in recent years there has been tremendous mathematical progress in this direction.[93–97] In the top-down methods, the details of the original mathematical representation of the wavefunction are not of prime importance and the function itself or a related function, e.g. a reduced density matrix or the electron density, is used as the input for the partitioning algorithm, yielding the AIM and their properties and interaction modes.[79–82] One of these methods, the *quantum theory of atoms in molecules* (QTAIM),[80,98,99] is considered in the next section.

## III. The QTAIM partitioning and its ramifications

The QTAIM partitioning scheme is based on the observation that the one-electron densities have a simple form in atoms with just a single maximum at the nucleus and then monotonically decaying away, while preserving these traits, to a large extent, in molecules.[80,100–105] The mentioned "robustness" of the atomic densities is rooted in the so-called "nearsightedness" of the electron densities that is observable in both computationally and experimentally derived densities,[106,107] but yet awaits a rigorous proof.[108–112] In fact, a pattern recognition algorithm,[113] may easily deduce the number and types of the constituent atoms of a molecule or a crystal from the electron densities since the amount of the electron density at the mentioned maximum is well-correlated with the atomic number of the nucleus (almost proportional to the third power of the nuclear charge).[80,114,115] The equation used for the partitioning of a molecule into AIM is the local zero-flux equation of the electron density: $\vec{\nabla}\rho_e(\vec{r}).\vec{n}(\vec{r}) = 0$, where: $\rho_e(\vec{r}) = N\int d\tau' \Psi_e^* \Psi_e$.[80] Note that $\Psi_e$ is the electronic wavefunction of the molecule while $N$ stands for the number of electrons and $d\tau'$ implies summing over spin variables of all electrons and integrating over spatial coordinates of all electrons except one arbitrary electron. The solutions of this equation are those surfaces where the unit vector normal to the surface, $\vec{n}(\vec{r})$, is perpendicular to the gradient of the electron density tangent to the surface and are called zero-flux surfaces.[80] Let us stress that the electronic wavefunctions are derived from the electronic Schrödinger equation,

$$\hat{H}_e \Psi_e(\{\vec{r}_i\};\{\vec{R}_\alpha\}) = E_e(\{\vec{R}_\alpha\}) \Psi_e(\{\vec{r}_i\};\{\vec{R}_\alpha\}), \quad \text{where:} \quad \hat{H}_e = \left(-\frac{1}{2}\right)\sum_i^N \nabla_i^2 + \sum_i^N \sum_{j>i}^N \frac{1}{|\vec{r}_i - \vec{r}_j|} - \sum_i^N \sum_\alpha^Q \frac{Z_\alpha}{|\vec{R}_\alpha - \vec{r}_i|}$$

(written in atomic units), implying that in this partitioning scheme the adiabatic approximation is assumed from the outset (the spin variables have been neglected for simplicity in this and subsequent representations of wavefunctions).[116] Accordingly, electrons are quantum particles in this Hamiltonian whereas nuclei are clamped point charges; there are $Q$ number of them each with its own charge $Z_\alpha$ and spatial position $\vec{R}_\alpha$, acting as the source of the external electric field for electrons. This is the reason that the electronic wavefunction and all the derived functions only depend parametrically on the nuclear spatial variables, $\{\vec{R}_\alpha\}$. The boundaries of AIM are those zero-flux surfaces that are not crossing the maxima of the electron density on the nuclei.[80] These boundaries partition the whole 3D space exhaustively into $Q$ number of 3D basins, $\bigcup_k^Q \Omega_k = R^3$, where each basin is conceived as an atom in a molecule.[80]

Let us stress that the zero-flux surfaces may sometimes encompass regions called "pseudo-atoms" which contain no clamped nucleus.[117–128] Pseudo-atoms rarely appear at the equilibrium geometries but their presence does not invalidate the following discussions and just increases the number of basins ($Q \to Q'$), though their chemical interpretation is not always straightforward (*vide infra*).[117–128] Finally, in contrast to some claims to the contrary,[80] the zero-flux surfaces have no significance from the viewpoint of the principles of QM.[129] Although these basins may fit the intuitional picture of AIM for many scientists,[80,98,99] the local zero-flux equation must be seen only as one of the auxiliary *ad hoc* principles of the QTAIM partitioning scheme as emphasized previously.

At the next step, it is desirable to derive the contribution of each atomic basin to the molecular expectation values, i.e. the molecular properties, and this is done by attributing a "property density" to each quantum observable: $A_e(\vec{r}) = \int d\tau' \text{Re}\left[\Psi_e^* \hat{A} \Psi_e\right]$.[80,111] Clearly, integration on the whole space yields the usual molecular expectation value, $\int_{R^3} d\vec{r} A_e(\vec{r}) = \langle \hat{A} \rangle$, which is a desirable feature. However, the property densities are not uniquely definable and it is feasible to construct an infinite number of property densities capable of yielding the corresponding expectation values, but differ locally: $A_e(\vec{r}) \neq A_e'(\vec{r})$, $\int_{R^3} d\vec{r} A_e(\vec{r}) = \int_{R^3} d\vec{r} A_e'(\vec{r}) = \langle \hat{A} \rangle$.[130,131] Like the case of AIM boundaries, this means that the general form of the integrant, $\text{Re}\left[\Psi_e^* \hat{A} \Psi_e\right]$, though a reasonable choice, is another auxiliary *ad hoc* principle in the QTAIM partitioning scheme. The contribution of an atomic basin to the molecular properties is straightforwardly derivable by the integration of each property density within its basin: $A_e(\Omega_k) = \int_{\Omega_k} d\vec{r} \, A_e(\vec{r})$. According to this definition the sum of all atomic contributions yields the molecular expectation value: $\langle \hat{A} \rangle = \sum_k^Q A_e(\Omega_k)$, and there is no contribution attributed to the inter-basin interactions.[80,111] In the case of the one-electron properties, $\hat{M} = \sum_{i=1}^N \hat{m}_i$, e.g. the kinetic energy, the property densities are also derivable directly from the spinless reduced first-order density matrix (1-RDM): $M_e(\vec{r}) = \hat{m}(\vec{r}) \rho_e^{(1)}(\vec{r}', \vec{r})\big|_{\vec{r}=\vec{r}'}$, where: $\rho_e^{(1)}(\vec{r}',\vec{r}) = N\int \tau'\Psi_e^*(\vec{r}_1',\vec{r}_2,...)\Psi_e(\vec{r}_1,\vec{r}_2,...)$.[80] Whereas, for the two-electron properties that do not include spatial derivatives, $\hat{G} = \sum_i^N \sum_{j>i}^N \hat{g}_{ij}$, e.g., the electron-electron interaction potential, the property densities are derivable from the spinless reduced second-order density matrix (2-RDM): $G_e(\vec{r}_1) = (1/2)\int d\vec{r}_2 \hat{g}(\vec{r}_1,\vec{r}_2) \rho_e^{(2)}(\vec{r}_1,\vec{r}_2)$, where: $\rho_e^{(2)}(\vec{r}_1,\vec{r}_2) = N(N-1)\int d\tau'' \Psi_e^* \Psi_e$. In the latter, the integration over $d\tau''$ implies summing over spin variables of all electrons except two arbitrary electrons denoted herein as particles "1" and "2". Another approach for the two-electron property partitioning has also been proposed that introduces inter-basin contributions and makes it possible to quantify the interactions of AIM: $\langle \hat{G} \rangle = \sum_k^Q G(\Omega_k) + \sum_k^Q \sum_{l>k}^Q G(\Omega_k,\Omega_l)$, where:

$$G(\Omega_k) = (1/2)\int_{\Omega_k} d\vec{r}_1 \int_{\Omega_k} d\vec{r}_2 \hat{g}(\vec{r}_1,\vec{r}_2) \rho_e^{(2)}(\vec{r}_1,\vec{r}_2)$$

and

$G(\Omega_k,\Omega_l) = \int_{\Omega_k} d\vec{r}_1 \int_{\Omega_l} d\vec{r}_2 \hat{g}(\vec{r}_1,\vec{r}_2) \rho_e^{(2)}(\vec{r}_1,\vec{r}_2)$.[132–137] Note that $\rho_e(\vec{r})$, $\rho_e^{(1)}(\vec{r}_1',\vec{r}_1)$ and $\rho_e^{(2)}(\vec{r}_1,\vec{r}_2)$ are all easily derivable from the non-diagonal form of the 2-RDM by simple integrations:

$\rho_e^{(2)}(\vec{r}_1',\vec{r}_2',\vec{r}_1,\vec{r}_2) = N(N-1)\int d\tau'' \Psi_e^*(\vec{r}_1',\vec{r}_2',\vec{r}_3...)\Psi_e(\vec{r}_1,\vec{r}_2,\vec{r}_3...)$.[138] So, the QTAIM partitioning can also be done if this function is directly computable instead of the electronic wavefunction.[139,140] Let us stress at this stage of development that there are molecular properties like the response properties that are not associated with any Hermitian operator, e.g. polarizability or magnetizability.[141] However, even for such properties if a proper definition of property density is introduced, their basin contribution may be computed through the previously introduced recipe (the interested reader may find the details in the original literature and particularly Keith's chapter in Ref. 99).[80,99,142] In the original formulation of the QTAIM,[80,111] there are other theoretical ingredients like the subsystem hypervirial theorem or the so-called variation derivation of the AIM.[80,111] Nevertheless, since these ingredients are of no practical use in the partitioning procedure, they are not considered herein. Based on the proposed definition of AIM, it may seem they are quite similar to the LEGO brick pieces and after a proper arrangement, the molecule and its properties are recovered.[143–145] However, this static "object-like" view of AIM is deceptive when taking into account that the number of electrons of each basin is fluctuating as will be demonstrated.

Let us start with the electron population of each basin. In order to derive the number of electrons, i.e. the electronic population, for each basin $\hat{A} = N\hat{1}$ is incorporated into the property density definition yielding: $N(\Omega_k) = \int_{\Omega_k} d\vec{r} \, \rho_e(\vec{r})$ and $\sum_k^Q N(\Omega_k) = N$, and the AIM charges are derivable from the basin electronic populations: $Q(\Omega_k) = Z_k - N(\Omega_k)$.[146] There is an alternative way of computing the electronic population by introducing the basin electron number distribution, which is the probability of $m$ number of electrons, $0 \leq m \leq N$, are in $\Omega_k$ while the rest are in the other basins:

$$P_m(\Omega_k) = \binom{N}{m} \int_{\Omega_k} d\vec{r}_1 \ldots \int_{\Omega_k} d\vec{r}_m \int_{R^3-\Omega_k} d\vec{r}_{m+1} \ldots \int_{R^3-\Omega_k} d\vec{r}_N \sum_{spins} \Psi_e^* \Psi_e \, .$$[80,147–149]

It is straightforward to demonstrate that the electronic population is the mean value of this distribution: $N(\Omega_k) = \sum_m^N m P_m(\Omega_k)$, where: $\sum_{m=0}^N P_m(\Omega_k) = 1$; the latter is the result of the wavefunction normalization.[80,147–149] Generally, for all $m$: $P_m(\Omega_k) > 0$, so the distribution has a non-zero variance and this opens up the door to compute the electron number fluctuation of the basin as the second statistical moment of the distribution:

$$Var(\Omega_k) = \sum_m^N (m - N(\Omega_k))^2 P_m(\Omega_k) = \sum_m^N m^2 P_m(\Omega_k) - \left[\sum_m^N m P_m(\Omega_k)\right]^2.$$[80,147,148] For a "closed" basin, i.e. with no electron exchange with surrounding basins, the distribution has just a single peak at one of $P_m(\Omega_k)$ values, let us call it $m'$, yielding: $P_{m'}(\Omega_k) = 1$ and $P_{m \neq m'}(\Omega_k) = 0$, and implying: $N(\Omega_k) = m'$ and $Var(\Omega_k) = 0$. Practically, this situation never happens for AIM at the equilibrium geometries and in general: $Var(\Omega_k) > 0$.[80,147–149] The variance can also be computed directly from the 2-RDM:

$$Var(\Omega_k) = \int_{\Omega_k} d\vec{r}_1 \int_{\Omega_k} d\vec{r}_2 \rho_e^{(2)}(\vec{r}_1, \vec{r}_2) + N(\Omega_k) - [N(\Omega_k)]^2,$$[80,147,148] revealing the fact that the fluctuation is not derivable just from the electron densities or the 1-RDM. Since the molecule is a closed system with respect to electron exchange, this implies: $Var(R^3) = 0$, and its partitioning into the basin and inter-basin contributions yields:

$$\sum_k^Q Var(\Omega_k) + 2 \sum_k^Q \sum_{l>k}^Q Cov(\Omega_k, \Omega_l) = 0.$$[150–152] In this expression, the inter-basin covariances appear naturally: $Cov(\Omega_k, \Omega_l) = \int_{\Omega_k} d\vec{r}_1 \int_{\Omega_l} d\vec{r}_2 \rho_e^{(2)}(\vec{r}_1, \vec{r}_2) - N(\Omega_k) N(\Omega_l),$[152] which are employed to introduce the so-called index of electronic delocalization: $\delta(\Omega_k, \Omega_l) = 2|Cov(\Omega_k, \Omega_l)|.$[150,151] $\delta(\Omega_k, \Omega_l)$ is gauging quantitatively the electron exchange between $\Omega_k$ and $\Omega_l$, and is usually interpreted chemically as a measure of bond orders between AIM.[153] Let us stress that recently there have been some attempts to generalize the idea of the basin electron number fluctuations to other one-electron properties as well.[154] All these demonstrate that an atom in a molecule not only interacts but also exchanges electrons and properties with surrounding AIM, and it is

better described as an "open atomic basin". Indeed, this image of AIM seems quite reasonable since the zero-flux surfaces are not physical but imaginary boundaries and electrons cannot sense them at all.

The last and probably, from the applied viewpoint, the most important ingredient of any partitioning scheme is its chemical interpretation. These are the "correspondence rules" making connections between chemical concepts, e.g. the covalency, iconicity, and aromaticity, and the properties and interaction modes of AIM. The algorithm of a partitioning scheme in itself does not reveal the correspondence rules, though it may sometimes offer some clues; the rules are most commonly derivable by employing the "inductive" reasoning. By applying the partitioning scheme to a set of reference molecules, one infers the rules and subsequently, they are tested on larger sets of molecules. All these may seem quite straightforward, but in practice, it is usually hard to find "universal" correspondence rules in a way that satisfies all scientists, and exceptions may always plague the promised universality. This is also the case for the QTAIM as there are growing examples that seemingly reasonable correspondence rules and their chemical interpretations have been challenged in subsequent studies.[155–161] A now classical example is the case of the so-called "hydrogen-hydrogen bonding" controversy,[162] which after two decades of intense studies has no generally accepted resolution yet.[163–199] However, let us stress that the QTAIM, by definition, cannot contradict the predictions of the underlying electronic Schrödinger equation, namely the resulting potential energy surface and various computed expectation values. If these predictions are in line with experimental data, the "story" emerging from the QTAIM partitioning, e.g. a proposal for a new type of bond or interaction, cannot be dismissed. Still, it seems that the ultimate merit of a story for practicing scientists is whether it can lead to discoveries or at least a reorganization of the knowledge in the field and some stories do not have such capacity. One must inevitably accept that the storytelling using the emergent entities is a tricky art and goes beyond employing some automated rules as it can be influenced by subjective judgments.[84] Such intricacies are also common in other scientific disciplines where the storytelling is part of the practice and there are usually competing stories that offer different causal narratives for a single system or phenomenon. Let us quote directly from Gould's brilliant easy on sociobiology: "When we examine the history of favored stories …, we do not trace a tale of increasing truth as one story replaces the last, but rather a chronicle of shifting fads and fashions".[200] Despite the storytelling aspects, the results of the QTAIM partitioning, i.e. the basin properties, can be used to develop new computational methods that are reconstructing the potential energy surfaces and molecular properties starting from AIM. Particularly interesting is a recent research program that tries to deduce the form of the energy terms and their associated atom-type parameters of force fields from the QTAIM partitioning to build reliable force fields for molecular simulations.[201–205] This program completes the logical steps between the lower and the intermediate levels since the phenomenological methods based on the concept of AIM were historically one of the driving forces behind the AIM-based partitioning developments, while they are currently benefiting from state of the art AIM partitioning methods.

## IV. Why the multi-component QTAIM?

The QTAIM partitioning scheme, as well as all other major partitioning algorithms,[79,82,81,206] have been designed to use the electronic wavefunctions as their inputs. For almost all usual applications in chemistry, this is a reasonable choice; however, some systems and phenomena are not governed by the electronic Schrödinger equation. In such cases, the adiabatic approximation is no longer applicable and the motion of certain nuclei may be coupled with the motion of electrons.[116,207] One method to handle such cases is to assume both electrons and certain nuclei as quantum particles while retaining enough clamped nuclei to avoid dealing with the total center of mass translational motion.[43,208,209] The resulting "multi-component" (MC) Schrödinger equation is solvable using various ab initio techniques developed recently and collectively called MC quantum chemistry.[210–216] The simplest case in this category is the two-component (TC) Schrödinger equation:

$$\hat{H}_{TC} = \left(-\frac{1}{2}\right)\sum_i^N \nabla_{e,i}^2 + \left(-\frac{1}{2m_2}\right)\sum_k^{N_2} \nabla_{2,k}^2 + \sum_i^N \sum_{j>i}^N \frac{1}{|\vec{r}_{e,i}-\vec{r}_{e,j}|} - \sum_i^N \sum_k^{N_2} \frac{q_2}{|\vec{r}_{e,i}-\vec{r}_{2,k}|} + \sum_k^{N_2} \sum_{l>k}^{N_2} \frac{q_2^2}{|\vec{r}_{2,k}-\vec{r}_{2,l}|} - \sum_i^N \sum_\alpha^Q \frac{Z_\alpha}{|\vec{R}_\alpha-\vec{r}_{e,i}|} + \sum_k^{N_2} \sum_\alpha^Q \frac{Z_\alpha q_2}{|\vec{R}_\alpha-\vec{r}_{2,k}|}$$

(written in atomic units), $\hat{H}_{TC}\Psi_{TC}\left(\{\vec{r}_{e,i}\},\{\vec{r}_{2,k}\};\{\vec{R}_\alpha\}\right) = E_{TC}\left(\{\vec{R}_\alpha\}\right)\Psi_{TC}\left(\{\vec{r}_{e,i}\},\{\vec{r}_{2,k}\};\{\vec{R}_\alpha\}\right)$. In this equation in addition to electrons and clamped nuclei, there is $N_2$ number of the second type of quantum particles, denoted by a subscript $2$, with mass $m_2$ and charge $q_2$ (the extension to the MC systems is straightforward and is not considered herein). Let us mention since the two types of quantum particles are "distinguishable", a single one-particle density cannot represent both quantum particles simultaneously as will be scrutinized in the next section. Another category of the MC quantum systems is the exotic systems where particles other than electrons and nuclei are present in the molecular system. The prime interest is muonic and positronic molecules where a single positively/negatively charged muon (hereafter called positive/negative muon for brevity) or a positron is added to the molecule opening a whole new field of the "exotic chemistry".[217–225] Since the mass ratios of positron and muons to electron, $\left(m_{positron}/m_{electron}\right)=1$ and $\left(m_{muon}/m_{electron}\right) \approx 206.768$,[226] are both much smaller than the same ratio in the case of nuclei, the adiabatic separation of the motion of electrons from these particles is no longer justified *per se*. Hence, assuming the nuclei as clamped point charges, the previously mentioned TC Schrödinger equation governs these systems as well, and the ab initio

MC quantum chemical techniques may be used to solve the equation. As reviewed recently,[154] the list of the MC quantum systems is much longer than the previously mentioned exotic species and contains also systems where the inter-particle interactions are not merely of Coulombic nature. This includes diverse sets of quantum systems across fundamental particle physics, nuclear physics, and the "truly" exotic species with no match in molecular physics and chemistry.[154] In such systems, there is clear evidence for an intermediate-level of organization and various forms of "internal clusterization" of the constituent particles may emerges very similar to the conventional AIM structures.[227–235] In fact, one may claim that the local zero-flux equation of the electron density is itself just a proper mathematical tool to decipher the electron clusterization in molecules around the nuclei. Naturally, the question arises whether one may extend the local zero-flux equation to the MC systems to grasp their clustered structures and then partition them into AIM-like basins. This is not a trivial task since all quantum particles may participate actively in the cluster formation and there is no reason to believe that just considering the one-particle density of one type of the constituent particles, e.g. electrons, is enough for a proper partitioning. In addition, a comprehensive partitioning scheme must be able to attribute the share of each basin to the total properties that now contains the contributions of all quantum particles and not just that of the electrons. As will be demonstrated in the next section, an extended version of the QTAIM, called MC-QTAIM, gathers all the required mathematical ingredients for a proper partitioning of the MC systems.

## V. How to design the MC-QTAIM?

The technical developments and certain applications of the MC-QTAIM have been detailed previously,[236–252] and the original motivations for its development have been reviewed elsewhere.[253] Herein, the historical path of the developments is not pursued and the MC-QTAIM is explained in a more axiomatic-like approach. As stressed in the previous section the MC wavefunctions contain the spatial variables of all types of the quantum particles: $\Psi_{MC}\left(\{\vec{r}_{1,1},...,\vec{r}_{1,N_1}\},...,\{\vec{r}_{n,1},...,\vec{r}_{n,N_n}\},...\{\vec{r}_{s,1},...,\vec{r}_{s,N_s}\};\{\vec{R}_\alpha\}\right)$. The one-particle densities are defined as follows: $\rho_n\left(\vec{r};\{\vec{R}_\alpha\}\right) = N_n \int d\tau'_n \Psi^*_{MC}\Psi_{MC}$, for each type of quantum particles where $N_n$ is the number of the particles of the $n-th$ type. Also, $d\tau'_n$ implies summing over spin variables of all quantum particles and integrating over spatial coordinates of all quantum particles except one arbitrary particle belonging to the $n-th$ type. Since there are $s$ types of distinguishable quantum particles, there are also $s$ number of distinct one-particle densities. Intuitively, one expects that a combination of all these one-particle densities to be incorporated into a zero-flux equation that delineates the boundaries of the atomic basins. One simple *ad hoc* suggestion is the product of these densities, which has been proposed and rejected immediately because of the evident mathematical inconsistencies some time ago.[254] Nonetheless, after a systematic and careful extension of the subsystem hypervirial theorem and the subsystem variation procedure for the MC systems,[242] which is not considered herein, the properly combined density was discovered.[242] The combined density was termed the *Gamma density*, which delivers the zero-flux surfaces that act as the AIM boundaries: $\vec{\nabla}\Gamma^{(s)}(\vec{r}).\vec{n}(\vec{r}) = 0$, $\Gamma^{(s)}(\vec{r}) = \sum_{n}^{s}(m_1/m_n)\rho_n(\vec{r})$. The Gamma density is an inversely mass-scaled sum of all the one-particle densities where the masses of the quantum particles, $m_1,...,m_s$, are ordered from the lightest to the heaviest. The electron density is the expectation value of the following operator: $\hat{\rho}_e(\vec{r}) = \sum_{i}^{N}\delta(\vec{r}_i-\vec{r})$,[255] and similarly, the corresponding operator for the Gamma density is: $\hat{\Gamma}^{(s)}(\vec{r}) = \sum_{n}^{s}(m_1/m_n)\sum_{i}^{N_n}\delta(\vec{r}_{n,i} - \vec{r})$, which reveals the nature of the combination.

The Gamma density has been used in various computational partitioning studies for several two-, three- and four-component molecular systems wherein the components were electrons, and positive particles with different masses like the hydrogen isotopes,[239,245,249] the positive muon,[244,247,248,251] the position,[237,252] and the negative muon.[244] In most of the considered systems containing positive particles, the local zero-flux equation of the Gamma density associates atomic basins to the hydrogen isotopes and the positive muon. Hence, the total number of basins, $P$, in these molecules is larger than the number of the clamped nuclei, $Q$. Each such basin contains a positive particle and surrounding electrons without containing any clamped nucleus and particularly, the muonic basin is the first example of an "exotic basin". This is easily quantifiable by extending the definition of the electron population to all types of quantum particles: $N_n(\Omega_k) = \int_{\Omega_k} d\vec{r}\, \rho_n(\vec{r})$ and $\sum_{n}^{s} N_n = N_{total} = \sum_{k}^{P}\sum_{n}^{s} N_n(\Omega_k)$.[240,243] Accordingly, the population of each positive particle is confined within a single atomic basin: $N_{n>1}(\Omega_l) \approx 1$ and $N_{n>1}(\Omega_{k \neq l}) \approx 0$. In contrast to the hydrogen isotopes and the positive muon, the positron in all studied cases was not capable of shaping its own atomic basin and always resided in one or rarely two of the basins associated with the clamped nuclei.[237] This reveals that the cluster formation capability of a positive particle depends on its mass and diminishes upon its decrease. A detailed computational study demonstrated that the "critical mass" at which a positive particle is capable of forming an atomic basin varies in various molecular systems and is tentatively somewhere between $50m_{electron}$ and $500m_{electron}$.[246] The amount of the electronic population within the positive particle basin further corroborates this picture since it is larger in the basins of the heavier particles.[246]

If the mass of a positive particle tends to infinity, the following non-trivial limit is derivable analytically: $\lim_{m_n \to \infty}(m_1/m_n)\rho_n(\vec{r}) \to 0$, $n > 1$, which cancels the positive particle's direct contribution to the Gamma density.[241] This is indeed the clamped particle limit and if all the positive particles are clamped in this manner, one arrives analytically at: $\lim_{m_2 \to \infty, \ldots, m_s \to \infty} \Gamma^{(s)}(\vec{r}) \to \rho_e(\vec{r})$.[241] This is a desirable feature of the MC-QTAIM scheme since it recovers the proper "single-component" limit where the QTAIM and corresponding zero-flux equation of the electron density works well. Based on these results, it is safe to conclude that at least in current state of affairs, the local zero-flux equation of the Gamma density seems to be a proper extension of the QTAIM's original partitioning scheme. Let us now introduce the basin properties within the context of the MC-QTAIM.

The property density and the basin contribution of the one-particle properties for the $n-th$ type of particles are: $M_n(\vec{r}) = \hat{m}_n(\vec{r})\rho_n^{(1)}(\vec{r}',\vec{r})\big|_{\vec{r}=\vec{r}'}$ and $M_n(\Omega_k) = \int_{\Omega_k} d\vec{r}\, M_n(\vec{r})$. Note that $\rho_n^{(1)}(\vec{r}',\vec{r}) = N_n \int d\tau_n' \Psi^*(\ldots,\vec{r}_{n,1}',\vec{r}_{n,2}\ldots\vec{r}_{n,N_n},\ldots)\Psi(\ldots,\vec{r}_{n,1},\vec{r}_{n,2},\ldots\vec{r}_{n,N_n},\ldots)$ is the generalized form of the 1-RDM for the $n-th$ type of particles where its diagonal element is the previously introduced one-particle density, $\rho_n(\vec{r})$.[241–243] The total contribution of each basin is the sum of the basin contributions of all quantum particles: $\tilde{M}(\Omega_k) = \sum_n^s M_n(\Omega_k)$, from which the molecular expectation value is recovered: $\langle \hat{M} \rangle = \left\langle \sum_n^s \sum_i^{N_n} \hat{m}_{n,i} \right\rangle = \sum_k^P \tilde{M}(\Omega_k)$. In the clamped nucleus limit, like the case of the one-particle densities discussed previously, one may demonstrate analytically: $\lim_{m_n \to \infty} M_n(\Omega_k) \to 0, \ n > 1$.[241] Consequently, in this limit, the basin contributions solely come from the electrons and the MC-QTAIM property partitioning *reduces* to that of the QTAIM.

Various applications of the property partitioning may be found in the original literature cited previously and let us just consider herein the concept of the *regional/basin positronic affinity* as an illustrative example.[237] The positron affinity (PA) of a molecule is the energy difference of the molecule before and after the addition of a positron.[223] In the case of a diatomic molecule, denoted as $AB$, it is computed as follows: $PA = E(AB) - E(AB, e^+)$. After partitioning the total energy into the basin contributions, the PA is also partitioned into basin contributions: $PA = \left(E_{AB}(\Omega_A) - E_{AB,e^+}(\Omega_A)\right) + \left(E_{AB}(\Omega_B) - E_{AB,e^+}(\Omega_B)\right) = PA(\Omega_A) + PA(\Omega_B)$.[237] According to the MC regional/basin virial theorem, when the molecule is at its equilibrium geometry, the total energy of each basin is equal to the basin's total kinetic energy, $\tilde{E}(\Omega_k) = -\tilde{T}(\Omega_k)$.[241,242] Thus, the basin PA is directly calculable from the basin kinetic energies:
$$PA(\Omega_k) = T^e_{AB,e^+}(\Omega_k) + T^p_{AB,e^+}(\Omega_k) - T^e_{AB}(\Omega_k),$$
where: $T^e(\Omega_k) = \int_{\Omega_k} d\vec{r} \left((-1/2)\nabla^2_{electron}\right) \rho^{(1)}_e(\vec{r}', \vec{r})\big|_{\vec{r}=\vec{r}'}, T^p(\Omega_k) = \int_{\Omega_k} d\vec{r} \left((-1/2)\nabla^2_{positron}\right) \rho^{(1)}_p(\vec{r}', \vec{r})\big|_{\vec{r}=\vec{r}'}$ (written in atomic units). The PAs of a set of diatomics including: $LiH$, $NaH$, $LiF$, $NaF$, $BeO$, $MgO$, $CN^-$ and $OH^-$ were computed at the MC-Hartree-Fock level and then partitioned according to the mentioned recipe.[237] The computed PAs of all these species are positive, however, the basin PAs are positive only for those basins containing the major share of the positron population. Accordingly, except $CN^-, e^+$, where the population of the positron is almost evenly distributed between the two basins, in the remaining positronic species the positron almost exclusively resides in the basin of the more electronegative atom making its PA positive. Whereas the other basin that does not contain a significant amount of the positron population has a negative PA and is destabilized relative to its corresponding basin in the purely electronic parent system. Thus, the computed basin PAs disclose a *mechanism* for the positron preferable "local" attachment to AIM and a concomitant stabilization that is hard to be revealed from mere examination of total PAs.

The two-particle properties that do not include the spatial derivatives are partitioned into the intra- and inter-basin contributions with the following recipe:

$$\langle \hat{G} \rangle = \left\langle \sum_n^s \left( \sum_i^{N_n} \sum_{j>i}^{N_n} \hat{g}_{n,ij} \right) + \sum_n^s \sum_{m>n}^s \left( \sum_i^{N_n} \sum_j^{N_m} \hat{g}_{nm,ij} \right) \right\rangle = \sum_k^P \tilde{G}(\Omega_k) + \sum_k^P \sum_{l>k}^P \tilde{G}(\Omega_k, \Omega_l),$$ where

$$\tilde{G}(\Omega_k) = \sum_n^s \left[ G_n(\Omega_k) + \sum_{m>n}^s G_{nm}(\Omega_k) \right]$$ and

$$\tilde{G}(\Omega_k, \Omega_l) = \sum_n^s \left[ G_n(\Omega_k, \Omega_l) + \sum_{m>n}^s G_{nm}(\Omega_k, \Omega_l) \right].^{154,249,252}$$ In these expressions:

$$G_n(\Omega_k) = (1/2) \int_{\Omega_k} d\vec{r}_1 \int_{\Omega_k} d\vec{r}_2 \ \hat{g}_{nn}(\vec{r}_1, \vec{r}_2) \rho_n^{(2)}(\vec{r}_1, \vec{r}_2), \quad G_{nm}(\Omega_k) = \int_{\Omega_k} d\vec{r}_n \int_{\Omega_k} d\vec{r}_m \ \hat{g}_{nm}(\vec{r}_n, \vec{r}_m) \rho_{nm}^{(2)}(\vec{r}_n, \vec{r}_m),$$

$$G_n(\Omega_k, \Omega_l) = \int_{\Omega_k} d\vec{r}_1 \int_{\Omega_l} d\vec{r}_2 \ \hat{g}_{nn}(\vec{r}_1, \vec{r}_2) \rho_n^{(2)}(\vec{r}_1, \vec{r}_2),$$

$$G_{nm}(\Omega_k, \Omega_l) = \int_{\Omega_k} d\vec{r}_n \int_{\Omega_l} d\vec{r}_m \ \hat{g}_{nm}(\vec{r}_n, \vec{r}_m) \rho_{nm}^{(2)}(\vec{r}_n, \vec{r}_m), \quad \text{where} \quad \rho_n^{(2)}(\vec{r}_1, \vec{r}_2) = N_n(N_n - 1) \int d\tau_n'' \Psi^* \Psi$$

and $\rho_{nm}^{(2)}(\vec{r}_n, \vec{r}_m) = N_n N_m \int d\tau_n''' \Psi^* \Psi$. Note that $d\tau_n''$ implies summing over spin variables of all quantum particles and integrating over spatial coordinates of all quantum particles except two arbitrary particles, denoted herein as particles "1" and "2", belonging to the $n-th$ type. Also, $d\tau_n'''$ implies summing over spin variables of all quantum particles and integrating over spatial coordinates of all quantum particles except two arbitrary particles; one belongs to the $n-th$ and the other to the $m-th$ types. In fact, $\rho_n^{(2)}(\vec{r}_1,\vec{r}_2)$ is the extension of the electronic 2-RDM for the $n-th$ type of particles while $\rho_{nm}^{(2)}(\vec{r}_n,\vec{r}_m)$ is a "pair density" originating from the distinguishability in particles of the two different types.[241–243] The most important application of this formalism is the partitioning of the two-particle interaction energy, which is an extended form of the *interacting quantum atoms* (IQA) energy partitioning scheme.[132–137] As an illustrative example let us consider the nature of the *positronic bond*,[256] as a completely novel type of bond,[257–260] where one or more positrons are acting as the glue between two anions. The MC-QTAIM partitioning of the simplest known system containing this type of bond, composed of two hydrides bonded by a single positron, $[H^-,e^+,H^-]$,[256] was done employing an ab initio MC-Hartree-Fock wavefunction.[252] The computational analysis generated two atomic basins each with a clamped proton and containing half of the electron and the positron populations imparting an even distribution of the total negative charge between the two basins. The partitioning of the electron-electron and the electron-positron interaction energies further revealed that the former has no stabilizing role in this species and even tries to keep the two hydrides apart. The main driving force behind the stabilization of this species was found to be the classical interaction of the positron density, which is centered between the two protons, and the electron density, as a component of the electron-positron interaction energy.[252] This is unparalleled with the two well-known main mechanisms of bond formation, i.e. covalent and ionic,[20,22,23,261–263] where the electrons are the bonding agents. The positronic bond is the first example of an "exotic bond" considered within the context of the MC-QTAIM as more examples awaiting the analysis.[264–266]

The last ingredient of the MC-QTAIM partitioning scheme comes from the quantification of the particle basin fluctuations by introducing the number distribution for each type of particles: $P_m^n(\Omega_k) = \binom{N_n}{m} \int_{\Omega_k} d\vec{r}_{n,1} \cdots \int_{\Omega_k} d\vec{r}_{n,m} \int_{R^3-\Omega_k} d\vec{r}_{n,m+1} \cdots \int_{R^3-\Omega_k} d\vec{r}_{n,N_n} \int d\omega_n \Psi^*\Psi$.[154,243] In this integration $d\omega_n$ implies summing over spin variables of all quantum particles and integrating over spatial coordinates of all quantum particles except the $n-th$ type. Each $P_m^n(\Omega_k)$ is the probability of observing $m$-particles from the $n-th$ type, $0 \le m \le N_n$, in the $k-th$ basin while the rest of $N_n - m$ particles are in the remaining basins: $R^3 - \Omega_k$. It is straightforward to demonstrate that the basin population of each type is the mean value of its associated number distribution: $N_n(\Omega_k) = \sum_m^{N_n} m P_m^n(\Omega_k)$, $\sum_{m=0}^{N_n} P_m^n(\Omega_k) = 1$.[154,243] The particle number fluctuation of each basin is the second statistical moment of the distribution:

$$Var_n(\Omega_k) = \sum_m^{N_n} m^2 P_m^n(\Omega_k) - \left[\sum_m^{N_n} m P_m^n(\Omega_k)\right]^2 = \int_{\Omega_k} d\vec{r}_1 \int_{\Omega_k} d\vec{r}_2 \rho_n^{(2)}(\vec{r}_1, \vec{r}_2) + N_n(\Omega_k) - [N_n(\Omega_k)]^2.$$[154,243]

For the heavy positive particles that may shape their basin and are confined within it, the distribution is simply: $P_1^n(\Omega_k) = 1$ and $P_{m \ne 1}^n(\Omega_k) = 0$, implying: $N_n(\Omega_k) = 1$ and $Var_n(\Omega_k) = 0$. The light positive particles nonetheless have the chance to fluctuate between basins as this is gauged by the covariance of the inter-basin contributions:

$$Var_n(R^3) = \sum_k^P Var_n(\Omega_k) + 2\sum_k^P \sum_{l>k}^P Cov_n(\Omega_k, \Omega_l) = 0, \quad Cov_n(\Omega_k, \Omega_l) = \int_{\Omega_k} d\vec{r}_1 \int_{\Omega_l} d\vec{r}_2 \rho_n^{(2)}(\vec{r}_1, \vec{r}_2) - N_n(\Omega_k)N_n(\Omega_l).$$[154,243]

An index of *particle delocalization* may be defined similarly to the previously defined electronic delocalization index as follows: $\delta_n(\Omega_k, \Omega_l) = 2|Cov_n(\Omega_k, \Omega_l)|$,[154,243] where one of its applications is for the antimatter molecules in which all electrons and nuclei are replaced by

their corresponding anti-particles. In these systems, the positrons are the bonding agents and the index quantifies the "anti-bond order".[154]

## VI. Conclusion and Prospects

The MC-QTAIM opens a completely new arena for the AIM partitioning of the MC systems. However, for the usual molecules composed of electrons and nuclei, except in rare cases,[253] the computational results are not far from those of the QTAIM.[239,245,249] For such cases the MC-QTAIM partitioning and the resulting AIM properties just reveal the subtle effects originating from the finite nuclear mass. For example, in the case of the hydrogen isotopes, the properties of a hydrogen basin containing a quantum proton do not differ seriously upon replacing quantum deuterium instead of the proton, which may be conceived as a kind of "isotope effect".[239,245,249] But, the genuinely interesting species for future applications of the MC-QTAIM are those composed of particles with small or no mass difference, which may form only "fragile" clusters. The prompt examples are systems composed of electrons and positrons, with or without clamped nuclei; sometimes it is even hard to unambiguously identify whether these species are atoms or molecules.[267–286]

There is also a whole plateau of "non-Coulombic" systems where the cluster formation is also taking place as the "nuclear molecules" are probably the most interesting examples.[227,228,233,287–295] In this class of nuclei, the protons and neutrons organize themselves as tightly bonded alpha clusters, i.e. Helium-4 nucleus, and the nucleus is modeled as a collection of these clusters.[296] The origin of this clustering is yet not fully understood,[297–299] however, accurate ab initio derived proton and neutron densities indeed reveal a clear pattern of cluster formation in real space.[300,301] The basic MC Hamiltonian of a nucleus, written based on the properties and interactions of protons and neutrons, apart from the Coulomb interactions between protons, contains distinct non-Coulombic nuclear interaction terms.[302–306] It is tempting to imagine a *quantum theory of clusters in nuclei* (QTCIN) that uses the associated TC wavefunction, containing the proton and neutron variables, as its input, and partitions the nuclear molecules into cluster basins. In fact, the discussed ingredients of the MC-QTAIM do not depend on the detailed nature of interactions between the constituent particles, and the same approach can be applied to diverse sets of quantum systems as far as the clusterization in real space forms the basis of the partitioning.[307,308]

Finally, let us stress that the developments of the MC-QTAIM may be seen as a way of extending the structural theory beyond the molecular realm to other areas of physics although the fruitfulness of such extension needs further scrutiny. Nevertheless, since the intermediate and high-level organizations are ubiquitous throughout the many-body systems at various scales of length and complexity,[235] there is a real demand for new theoretical developments to understand these types of organizations.

## Acknowledgments

The author is grateful to Mohammad Goli, Paul Popelier and Cina Foroutan-Nejad for their constructive comments on a previous draft.

## References

[1] W. Poundstone, *The Recursive Universe: Cosmic Complexity and the Limits of Scientific Knowledge*, American First edition (Contemporary Books, Chicago, 1985).
[2] S.J. Blundell, Philosophica **92**, 139 (2017).
[3] A.J. Rocke, *The Quiet Revolution: Hermann Kolbe and the Science of Organic Chemistry* (University of California Press, Berkeley, 1993).
[4] A.J. Rocke, *Chemical Atomism in the Nineteenth Century: From Dalton to Cannizzaro* (Ohio State University Press, Columbus, 1984).
[5] J.J. Sylvester, Nature **17**, 284 (1878).
[6] N. Trinajstic, *Chemical Graph Theory*, 2 edition (CRC Press, Boca Raton, 1992).
[7] D.H. Rouvray, J. Mol. Struct.: THEOCHEM **336**, 101 (1995).

[8] P.J. Ramberg, *Chemical Structure, Spatial Arrangement: The Early History of Stereochemistry, 1874–1914* (Routledge, 2017).
[9] C.K. Ingold, Nature **133**, 946 (1934).
[10] C.A. Russell, *The History of Valency* (Leicester University Press, Leicester, 1971).
[11] M.D. Saltzman, J. Chem. Educ. **57**, 484 (1980).
[12] M.D. Saltzman, Nat. Prod. Rep. **4**, 53 (1987).
[13] G.N. Lewis, J. Am. Chem. Soc. **38**, 762 (1916).
[14] I. Langmuir, J. Am. Chem. Soc. **41**, 868 (1919).
[15] G.N. Lewis, *Valence and the Structure of Atoms and Molecules* (The Chemical Catalog Company, 1923).
[16] L. Zhao, W.H.E. Schwarz, and G. Frenking, Nat. Rev. Chem. **3**, 35 (2019).
[17] H. Hettema, editor , *Quantum Chemistry: Classic Scientific Papers* (World Scientific Pub Co Inc, Singapore ; River Edge, NJ ; London, 2000).
[18] T. Helgaker, P. Jørgensen, and J. Olsen, *Molecular Electronic-Structure Theory* (John Wiley & Sons, Ltd, Chichester, UK, 2000).
[19] R.B. Laughlin and D. Pines, PNAS **97**, 28 (2000).
[20] L. Pauling, *The Nature of the Chemical Bond and the Structure of Molecules and Crystals: An Introduction to Modern Structural Chemistry*, Third edition (Cornell University Press, Ithaca, New York, 1960).
[21] P. Ball, Nature **469**, 26 (2011).
[22] G. Frenking and S. Shaik, editors , *The Chemical Bond: Fundamental Aspects of Chemical Bonding*, 1st edition (Wiley-VCH, 2014).
[23] G. Frenking and S. Shaik, editors , *The Chemical Bond: Chemical Bonding Across the Periodic Table*, 1st edition (Wiley-VCH, Weinheim, 2014).
[24] A. Neugebauer and G. Häfelinger, J. Phys. Org. Chem. **15**, 677 (2002).
[25] F.A. Carey and R.J. Sundberg, *Advanced Organic Chemistry, Part A: Structure and Mechanisms*, 5th edition (Springer, New York, 2008).
[26] H. Schnöckel, Chem. Rev. **110**, 4125 (2010).
[27] P. Jena, J. Phys. Chem. Lett. **4**, 1432 (2013).
[28] Z. Luo and A.W. Castleman, Acc. Chem. Res. **47**, 2931 (2014).
[29] A. Pinkard, A.M. Champsaur, and X. Roy, Acc. Chem. Res. **51**, 919 (2018).
[30] E.A. Doud, A. Voevodin, T.J. Hochuli, A.M. Champsaur, C. Nuckolls, and X. Roy, Nat Rev Mater **5**, 371 (2020).
[31] P. Jena and Q. Sun, editors , *Superatoms: Principles, Synthesis and Applications*, 1st edition (Wiley, Hoboken, NJ, 2021).
[32] S.S. Shaik and P.C. Hiberty, *A Chemist's Guide to Valence Bond Theory*, 1st edition (Wiley-Interscience, 2010).
[33] T.A. Albright, J.K. Burdett, and M.-H. Whangbo, *Orbital Interactions in Chemistry*, 2nd edition (Wiley-Interscience, Hoboken, New Jersey, 2013).
[34] N.T. Anh, *Frontier Orbitals: A Practical Manual* (Wiley, Chichester, England ; Hoboken, NJ, 2007).
[35] I. Fleming, *Molecular Orbitals and Organic Chemical Reactions*, Student edition (Wiley, Chichester, West Sussex, U.K, 2009).
[36] E.V. Anslyn and D.A. Dougherty, *Modern Physical Organic Chemistry*, illustrated edition edition (University Science, Sausalito, CA, 2005).
[37] P. Karen, Angew. Chem. Int. Ed. **54**, 4716 (2015).

[38] S. Fliszar, *Atomic Charges, Bond Properties, and Molecular Energies*, 1st edition (Wiley-Interscience, Hoboken, N.J, 2008).

[39] A.J. Stone, *The Theory of Intermolecular Forces*, second edition (Clarendon Press, oxford, 2013).

[40] R.G. Woolley, Adv. Phys. **25**, 27 (1976).

[41] H. Primas, *Chemistry, Quantum Mechanics and Reductionism: Perspectives in Theoretical Chemistry* (Springer Berlin Heidelberg, Berlin, Heidelberg, 1983).

[42] H. Primas, Acta Polytech. Scand., Ma, Math. Comput. **91**, 83 (1998).

[43] B.T. Sutcliffe and R.G. Woolley, Phys. Chem. Chem. Phys. **7**, 3664 (2005).

[44] P.W. Anderson, Science **177**, 393 (1972).

[45] I. Prigogine, *From Being to Becoming: Time and Complexity in the Physical Sciences*, First Edition (W H Freeman & Co, San Francisco, 1981).

[46] S.A. Kauffman, *The Origins of Order: Self-Organization and Selection in Evolution*, 1 edition (Oxford University Press, U.S.A., New York, 1993).

[47] J.-M. Lehn, *Supramolecular Chemistry: Concepts and Perspectives*, 1 edition (Wiley-VCH, Weinheim ; New York, 1995).

[48] S. Kauffman, *At Home in the Universe: The Search for the Laws of Self-Organization and Complexity*, Revised ed. edition (Oxford University Press, New York, 1996).

[49] P. Bak, *How Nature Works: The Science of Self-Organized Criticality*, First Softcover edition (Copernicus, New York, 1999).

[50] S. Johnson, *Emergence: The Connected Lives of Ants, Brains, Cities, and Software*, Reprint edition (Scribner, New York, NY, 2002).

[51] A. Huttemann, *What's Wrong With Microphysicalism?*, 1st edition (Routledge, 2004).

[52] R.B. Laughlin, *A Different Universe: Reinventing Physics from the Bottom Down*, New Ed edition (Basic Books, New York, NY, 2006).

[53] R.W. Batterman, *The Devil in the Details: Asymptotic Reasoning in Explanation, Reduction, and Emergence*, 1 edition (Oxford University Press, Oxford, 2006).

[54] E. Mayr, *What Makes Biology Unique?: Considerations on the Autonomy of a Scientific Discipline*, 1 edition (Cambridge University Press, London, 2007).

[55] M.A. Bedau and P. Humphreys, editors , *Emergence: Contemporary Readings in Philosophy and Science*, Annotated edition (A Bradford Book, Cambridge, Mass, 2008).

[56] P.W. Anderson, *More And Different: Notes From A Thoughtful Curmudgeon* (World Scientific, 2011).

[57] J.-M. Lehn, Angew. Chem. Int. Ed. **52**, 2836 (2013).

[58] S. Chibbaro, L. Rondoni, and A. Vulpiani, *Reductionism, Emergence and Levels of Reality: The Importance of Being Borderline*, 2014th edition (Springer, New York, 2014).

[59] M. Morrison, *Reconstructing Reality: Models, Mathematics, and Simulations*, 1st edition (Oxford University Press, 2014).

[60] B. Falkenburg and M. Morrison, editors , *Why More Is Different: Philosophical Issues in Condensed Matter Physics and Complex Systems* (Springer-Verlag, Berlin Heidelberg, 2015).

[61] S.A. Kauffman, *Humanity in a Creative Universe*, 1 edition (Oxford University Press, New York, NY, 2016).

[62] P. Humphreys, *Emergence: A Philosophical Account* (Oxford University Press, 2016).

[63] G. Ellis, *How Can Physics Underlie the Mind?: Top-Down Causation in the Human Context* (Springer-Verlag, Berlin Heidelberg, 2016).

[64] M.P. Paoletti and F. Orilia, editors, *Philosophical and Scientific Perspectives on Downward Causation*, 1 edition (Routledge, New York, 2017).

[65] S.A. Kauffman, *A World Beyond Physics: The Emergence and Evolution of Life*, 1st edition (Oxford University Press, New York, NY, 2019).

[66] S. Gibb, R.F. Hendry, and T. Lancaster, editors, *The Routledge Handbook of Emergence*, 1 edition (Routledge, New York, 2019).

[67] S. Weinberg, *Dreams of a Final Theory: The Scientist's Search for the Ultimate Laws of Nature*, Reprint edition (Vintage, New York, 1994).

[68] R.D. Mattuck, *A Guide to Feynman Diagrams in the Many-Body Problem: Second Edition*, 2 edition (Dover Publications, New York, 1992).

[69] D. McComb, *Renormalization Methods: A Guide For Beginners*, 1 edition (Oxford University Press, USA, Oxford, 2008).

[70] T. Lancaster and S.J. Blundell, *Quantum Field Theory for the Gifted Amateur*, 1 edition (Oxford University Press, Oxford, 2014).

[71] U. Burkert and N.L. Allinger, *Molecular Mechanics* (American Chemical Society, Washington, DC, 1982).

[72] A.K. Rappe and C.J. Casewit, *Molecular Mechanics Across Chemistry* (University Science Books, Sausalito, Calif, 1997).

[73] N.L. Allinger and D.W. Rogers, *Molecular Structure: Understanding Steric and Electronic Effects from Molecular Mechanics*, 1 edition (Wiley, Hoboken, N.J, 2010).

[74] M. Finnis, *Interatomic Forces in Condensed Matter* (Oxford University Press, Oxford, 2010).

[75] A.M. Pendás and E. Francisco, J. Chem. Theory Comput. **15**, 1079 (2019).

[76] S. Haroche and J.-M. Raimond, *Exploring the Quantum: Atoms, Cavities, and Photons*, 1st edition (OUP Oxford, 2006).

[77] H.-P. Breuer, *The Theory of Open Quantum Systems* (Oxford University Press, USA, Oxford, 2007).

[78] M.A. Schlosshauer, *Decoherence and the Quantum-to-Classical Transition* (Springer, Berlin ; London, 2008).

[79] R.G. Parr and W. Yang, *Density-Functional Theory of Atoms and Molecules* (Oxford University Press USA, New York, NY, 1994).

[80] R.F.W. Bader, *Atoms in Molecules: A Quantum Theory* (Clarendon Press, 1994).

[81] F. Weinhold and C.R. Landis, *Valency and Bonding: A Natural Bond Orbital Donor-Acceptor Perspective*, 1 edition (Cambridge University Press, Cambridge, UK ; New York, 2005).

[82] I. Mayer, *Bond Orders and Energy Components: Extracting Chemical Information from Molecular Wave Functions* (CRC Press, Boca Raton, 2016).

[83] B.T. Sutcliffe, Int. J. Quantum Chem. **58**, 645 (1996).

[84] W.H.E. Schwarz and H. Schmidbaur, Chem. Eur. J. **18**, 4470 (2012).

[85] P.L. Ayers, R.J. Boyd, P. Bultinck, M. Caffarel, R. Carbó-Dorca, M. Causá, J. Cioslowski, J. Contreras-Garcia, D.L. Cooper, P. Coppens, C. Gatti, S. Grabowsky, P. Lazzeretti, P. Macchi, Á. Martín Pendás, P.L.A. Popelier, K. Ruedenberg, H. Rzepa, A. Savin, A. Sax, W.H.E. Schwarz, S. Shahbazian, B. Silvi, M. Solà, and V. Tsirelson, Comput. Theor. Chem. **1053**, 2 (2015).

[86] J. Andrés, P.W. Ayers, R.A. Boto, R. Carbó-Dorca, H. Chermette, J. Cioslowski, J. Contreras-García, D.L. Cooper, G. Frenking, C. Gatti, F. Heidar-Zadeh, L. Joubert, Á.M.

Pendás, E. Matito, I. Mayer, A.J. Misquitta, Y. Mo, J. Pilmé, P.L.A. Popelier, M. Rahm, E. Ramos-Cordoba, P. Salvador, W.H.E. Schwarz, S. Shahbazian, B. Silvi, M. Solà, K. Szalewicz, V. Tognetti, F. Weinhold, and É.-L. Zins, J. Comput. Chem. **40**, 2248 (2019).

[87] Moffitt W. and R. E. Keightley, Proc. R. Soc. A: Math. Phys. Eng. Sci. **210**, 245 (1951).
[88] W. Moffitt, Rep. Prog. Phys. **17**, 173 (1954).
[89] T. Arai, Rev. Mod. Phys. **32**, 370 (1960).
[90] R. McWeeny, in *Quantum Systems in Chemistry and Physics. Trends in Methods and Applications*, edited by R. McWeeny, J. Maruani, Y.G. Smeyers, and S. Wilson (Springer Netherlands, Dordrecht, 1997), pp. 7–26.
[91] J.C. Tully, Adv. Chem. Phys. **42**, 63 (2007).
[92] P.W. Langhoff, J. Phys. Chem. **100**, 2974 (1996).
[93] J.D. Mills, J.A. Boatz, and P.W. Langhoff, Phys. Rev. A **98**, 012506 (2018).
[94] P.W. Langhoff, J.A. Boatz, R.J. Hinde, and J.A. Sheehy, J. Chem. Phys. **121**, 9323 (2004).
[95] P.W. Langhoff, R.J. Hinde, J.D. Mills, and J.A. Boatz, Theor. Chem. Acc. **120**, 199 (2008).
[96] M. Ben-Nun, J.D. Mills, R.J. Hinde, C.L. Winstead, J.A. Boatz, G.A. Gallup, and P.W. Langhoff, J. Phys. Chem. A **113**, 7687 (2009).
[97] P.W. Langhoff, J.D. Mills, and J.A. Boatz, J. Math. Phys. **59**, 072105 (2018).
[98] P.L. Popelier, *Atoms in Molecules: An Introduction*, 1 edition (Prentice Hall, Harlow, 2000).
[99] C.F. Matta and R.J. Boyd, *The Quantum Theory of Atoms in Molecules: From Solid State to DNA and Drug Design* (Wiley-VCH, Weinheim, 2007).
[100] H. Weinstein, P. Politzer, and S. Srebrenik, Theoret. Chim. Acta **38**, 159 (1975).
[101] J.C. Angulo, R.J. Yáñez, J.S. Dehesa, and E. Romera, Int. J. Quantum Chem. **58**, 11 (1996).
[102] P.W. Ayers and R.G. Parr, Int. J. Quantum Chem. **95**, 877 (2003).
[103] Z.A. Keyvani, S. Shahbazian, and M. Zahedi, Chem. Eur. J. **22**, 5003 (2016).
[104] Z.A. Keyvani, S. Shahbazian, and M. Zahedi, ChemPhysChem **17**, 3260 (2016).
[105] S. Shahbazian, Int. J. Quantum Chem. **118**, e25637 (2018).
[106] P. Coppens, *X-Ray Charge Densities and Chemical Bonding* (International Union of Crystallography, Chester, England : Oxford ; New York, 1997).
[107] C. Gatti and P. Macchi, editors, *Modern Charge-Density Analysis* (Springer Netherlands, 2012).
[108] W. Kohn and A. Yaniv, PNAS **75**, 5270 (1978).
[109] E. Prodan and W. Kohn, PNAS **102**, 11635 (2005).
[110] E. Prodan, Phys. Rev. B **73**, 085108 (2006).
[111] R.F.W. Bader, J. Phys. Chem. A **111**, 7966 (2007).
[112] S. Fias, F. Heidar-Zadeh, P. Geerlings, and P.W. Ayers, PNAS **114**, 11633 (2017).
[113] R.O. Duda, P.E. Hart, and D.G. Stork, *Pattern Classification*, 2nd edition (Wiley-Interscience, New York, 2000).
[114] A.M. Simas, R.P. Sagar, A.C.T. Ku, and V.H. Smith Jr., Can. J. Chem. **66**, 1923 (1988).
[115] A. Marefat Khah and S. Shahbazian, ArXiv:1510.07299 (2015).
[116] H. Mustroph, ChemPhysChem **17**, 2616 (2016).
[117] W.L. Cao, C. Gatti, P.J. MacDougall, and R.F.W. Bader, Chem. Phys. Lett. **141**, 380 (1987).
[118] C. Gatti, P. Fantucci, and G. Pacchioni, Theoret. Chim. Acta **72**, 433 (1987).
[119] J. Cioslowski, J. Phys. Chem. **94**, 5496 (1990).
[120] K.E. Edgecombe, R.O. Esquivel, V.H. Smith, and F. Müller-Plathe, J. Chem. Phys. **97**, 2593 (1992).

[121] K.E. Edgecombe, J. Vedene H. Smith, and F. Müller-Plathe, Z. Naturforsch. **48a**, 127 (1993).
[122] G.I. Bersuker, C. Peng, and J.E. Boggs, J. Phys. Chem. **97**, 9323 (1993).
[123] A.M. Pendás, M.A. Blanco, A. Costales, P.M. Sánchez, and V. Luaña, Phys. Rev. Lett. **83**, 1930 (1999).
[124] P. Mori-Sánchez, A.M. Pendás, and V. Luaña, Phys. Rev. B **63**, 125103 (2001).
[125] G.K.H. Madsen, P. Blaha, and K. Schwarz, J. Chem. Phys. **117**, 8030 (2002).
[126] V. Luaña, P. Mori-Sánchez, A. Costales, M.A. Blanco, and A.M. Pendás, J. Chem. Phys. **119**, 6341 (2003).
[127] A. Costales, M.A. Blanco, A. Martín Pendás, P. Mori-Sánchez, and V. Luaña, J. Phys. Chem. A **108**, 2794 (2004).
[128] L.A. Terrabuio, T.Q. Teodoro, M.G. Rachid, and R.L.A. Haiduke, J. Phys. Chem. A **117**, 10489 (2013).
[129] S. Shahbazian, Int. J. Quantum Chem. **111**, 4497 (2011).
[130] D.B. Cook, *Schrodinger's Mechanics* (World Scientific Publishing Company, Singapore ; Teaneck, N.J, 1989).
[131] J.S.M. Anderson, P.W. Ayers, and J.I.R. Hernandez, J. Phys. Chem. A **114**, 8884 (2010).
[132] P.L.A. Popelier and D.S. Kosov, J. Chem. Phys. **114**, 6539 (2001).
[133] M.A. Blanco, A. Martín Pendás, and E. Francisco, J. Chem. Theory Comput. **1**, 1096 (2005).
[134] E. Francisco, A. Martín Pendás, and M.A. Blanco, J. Chem. Theory Comput. **2**, 90 (2006).
[135] M. García-Revilla, E. Francisco, P.L.A. Popelier, and A.M. Pendás, ChemPhysChem **14**, 1211 (2013).
[136] J.M. Guevara-Vela, E. Francisco, T. Rocha-Rinza, and Á. Martín Pendás, Molecules **25**, 4028 (2020).
[137] A.F. Silva, L.J. Duarte, and P.L.A. Popelier, Struct. Chem. **31**, 507 (2020).
[138] R. McWeeny, *Methods of Molecular Quantum Mechanics*, 2nd edition (Academic Press, London, 1992).
[139] D.A. Mazziotti, Acc. Chem. Res. **39**, 207 (2006).
[140] D.A. Mazziotti, Chem. Rev. **112**, 244 (2012).
[141] P.W. Atkins and R.S. Friedman, *Molecular Quantum Mechanics*, 5th edition (Oxford University Press, Oxford ; New York, 2010).
[142] R.F.W. Bader, Mol. Phys. **100**, 3333 (2002).
[143] J.P. Mathias and J.F. Stoddart, Chem. Soc. Rev. **21**, 215 (1992).
[144] C.V.K. Sharma, J. Chem. Educ. **78**, 617 (2001).
[145] C. Schafmeister, Sci. Am. **296 (Feb.)**, 76 (2007).
[146] R.F.W. Bader and C.F. Matta, J. Phys. Chem. A **108**, 8385 (2004).
[147] R.F.W. Bader and M.E. Stephens, Chem. Phys. Lett. **26**, 445 (1974).
[148] R.F.W. Bader and M.E. Stephens, J. Am. Chem. Soc. **97**, 7391 (1975).
[149] Á.M. Pendás and E. Francisco, ChemPhysChem **20**, 2722 (2019).
[150] X. Fradera, M.A. Austen, and R.F.W. Bader, J. Phys. Chem. A **103**, 304 (1999).
[151] X. Fradera, J. Poater, S. Simon, M. Duran, and M. Solà, Theor. Chem. Acc. **108**, 214 (2002).
[152] E. Francisco, A. Martín Pendás, and M.A. Blanco, J. Chem. Phys. **126**, 094102 (2007).
[153] C. Outeiral, M.A. Vincent, Á.M. Pendás, and P.L.A. Popelier, Chem. Sci. **9**, 5517 (2018).
[154] S. Shahbazian, ArXiv:2109.15008 (2021).

[155] L.J. Farrugia, C. Evans, and M. Tegel, J. Phys. Chem. A **110**, 7952 (2006).
[156] S. Shahbazian, ArXiv:1306.6350 (2013).
[157] C. Foroutan-Nejad, S. Shahbazian, and R. Marek, Chem. Eur. J. **20**, 10140 (2014).
[158] S. Shahbazian, Chem. Eur. J. **24**, 5401 (2018).
[159] C.R. Wick and T. Clark, J. Mol. Model. **24**, 142 (2018).
[160] M. Jabłoński, J. Comput. Chem. **39**, 2183 (2018).
[161] M. Jabłoński, ChemistryOpen **8**, 497 (2019).
[162] C.F. Matta, J. Hernández-Trujillo, T.-H. Tang, and R.F.W. Bader, Chem. Eur. J. **9**, 1940 (2003).
[163] J. Cioslowski, S.T. Mixon, and W.D. Edwards, J. Am. Chem. Soc. **113**, 1083 (1991).
[164] J. Cioslowski and S.T. Mixon, J. Am. Chem. Soc. **114**, 4382 (1992).
[165] J. Cioslowski and S.T. Mixon, Can. J. Chem. **70**, 443 (1992).
[166] C.-C. Wang, T.-H. Tang, L.-C. Wu, and Y. Wang, Acta Cryst. A **60**, 488 (2004).
[167] J. Poater, M. Solà, and F.M. Bickelhaupt, Chem. Eur. J. **12**, 2889 (2006).
[168] R.F.W. Bader, Chem. Eur. J. **12**, 2896 (2006).
[169] C.F. Matta, in *Hydrogen Bonding—New Insights*, edited by S.J. Grabowski (Springer Netherlands, Dordrecht, 2006), pp. 337–375.
[170] L.F. Pacios and L. Gómez, Chem. Phys. Lett. **432**, 414 (2006).
[171] J. Hernández-Trujillo and C.F. Matta, Struct Chem **18**, 849 (2007).
[172] J. Poater, R. Visser, M. Solà, and F.M. Bickelhaupt, J. Org. Chem. **72**, 1134 (2007).
[173] L.F. Pacios, Struct Chem **18**, 785 (2007).
[174] J. Poater, J.J. Dannenberg, M. Sola, and F.M. Bickelhaupt, Int. J. Chem. Model. **1**, 63 (2008).
[175] S. Grimme, C. Mück-Lichtenfeld, G. Erker, G. Kehr, H. Wang, H. Beckers, and H. Willner, Angew. Chem. Int. Ed. **48**, 2592 (2009).
[176] H. Jacobsen, Dalton Trans. **39**, 5426 (2010).
[177] J. Poater, M. Sola, and F.M. Bickelhaupt, Int. J. Chem. Model. **2**, 181 (2010).
[178] I. Cukrowski and C.F. Matta, Chem. Phys. Lett. **499**, 66 (2010).
[179] I. Cukrowski, K.K. Govender, M.P. Mitoraj, and M. Srebro, J. Phys. Chem. A **115**, 12746 (2011).
[180] J. Echeverría, G. Aullón, D. Danovich, S. Shaik, and S. Alvarez, Nat. Chem. **3**, 323 (2011).
[181] G.V. Baryshnikov, B.F. Minaev, and V.A. Minaeva, J. Struct. Chem. **52**, 1051 (2011).
[182] R.D. Hancock and I.V. Nikolayenko, J. Phys. Chem. A **116**, 8572 (2012).
[183] A.V. Vashchenko and T.N. Borodina, J. Struct. Chem. **54**, 479 (2013).
[184] F. Cortés-Guzmán, J. Hernández-Trujillo, and G. Cuevas, J. Phys. Chem. A **107**, 9253 (2003).
[185] F. Weinhold, P. von R. Schleyer, and W.C. McKee, J. Comput. Chem. **35**, 1499 (2014).
[186] R. Kalescky, E. Kraka, and D. Cremer, J. Phys. Chem. A **118**, 223 (2014).
[187] D.A. Safin, M.G. Babashkina, K. Robeyns, M.P. Mitoraj, P. Kubisiak, and Y. Garcia, Chem. Eur. J. **21**, 16679 (2015).
[188] S. Jenkins, J.R. Maza, T. Xu, D. Jiajun, and S.R. Kirk, J. Comput. Chem. **115**, 1678 (2015).
[189] C.F. Matta, S.A. Sadjadi, D.A. Braden, and G. Frenking, J. Comput. Chem. **37**, 143 (2016).
[190] D. Jiajun, Y. Xu, T. Xu, R. Momen, S.R. Kirk, and S. Jenkins, Chem. Phys. Lett. **651**, 251 (2016).
[191] J. Poater, J. Paauwe, S. Pan, G. Merino, C.F. Guerra, and F.M. Bickelhaupt, Mol. Astrophys. **8**, 19 (2017).

[192] D. Myburgh, S. von Berg, and J. Dillen, J. Comput. Chem. **39**, 2273 (2018).
[193] P.L.A. Popelier, P.I. Maxwell, J.C.R. Thacker, and I. Alkorta, Theor. Chem. Acc. **138**, 12 (2018).
[194] J.S. Lomas, Magn. Reson. Chem. **57**, 1121 (2019).
[195] R. Laplaza, R.A. Boto, J. Contreras-García, and M.M. Montero-Campillo, Phys. Chem. Chem. Phys. **22**, 21251 (2020).
[196] M.P. Mitoraj, F. Sagan, D.W. Szczepanik, J.H. de Lange, A.L. Ptaszek, D.M.E. van Niekerk, and I. Cukrowski, ChemPhysChem **21**, 494 (2020).
[197] M.-L.Y. Riu, G. Bistoni, and C.C. Cummins, J. Phys. Chem. A **125**, 6151 (2021).
[198] T.G. Bates, J.H. de Lange, and I. Cukrowski, J. Comput. Chem. **42**, 706 (2021).
[199] B. Landeros-Rivera and J. Hernández-Trujillo, ChemPlusChem **87**, e202100492 (2022).
[200] S.J. Gould, New Scientist **80**, 530 (1978).
[201] P.L.A. Popelier and F.M. Aicken, J. Am. Chem. Soc. **125**, 1284 (2003).
[202] P.L.A. Popelier and F.M. Aicken, Chem. Eur. J. **9**, 1207 (2003).
[203] P.L.A. Popelier, Int. J. Quantum Chem. **115**, 1005 (2015).
[204] P.L.A. Popelier, Phys. Scr. **91**, 033007 (2016).
[205] B.C.B. Symons, M.K. Bane, and P.L.A. Popelier, J. Chem. Theory Comput. **17**, 7043 (2021).
[206] A. Savin, R. Nesper, S. Wengert, and T.F. Fässler, Angew. Chem. Int. Ed. **36**, 1808 (1997).
[207] M. Baer, *Beyond Born-Oppenheimer: Electronic Nonadiabatic Coupling Terms and Conical Intersections*, 1 edition (Wiley-Interscience, Hoboken, N.J, 2006).
[208] S. Bubin, M. Pavanello, W.-C. Tung, K.L. Sharkey, and L. Adamowicz, Chem. Rev. **113**, 36 (2013).
[209] J. Mitroy, S. Bubin, W. Horiuchi, Y. Suzuki, L. Adamowicz, W. Cencek, K. Szalewicz, J. Komasa, D. Blume, and K. Varga, Rev. Mod. Phys. **85**, 693 (2013).
[210] M. Tachikawa, K. Mori, H. Nakai, and K. Iguchi, Chem. Phys. Lett. **290**, 437 (1998).
[211] S.P. Webb, T. Iordanov, and S. Hammes-Schiffer, J. Chem. Phys. **117**, 4106 (2002).
[212] H. Nakai, Int. J. Quantum Chem. **107**, 2849 (2007).
[213] T. Ishimoto, M. Tachikawa, and U. Nagashima, Int. J. Quantum Chem. **109**, 2677 (2018).
[214] A. Reyes, F. Moncada, and J. Charry, Int. J. Quantum Chem. **119**, e25705 (2019).
[215] F. Pavošević, T. Culpitt, and S. Hammes-Schiffer, Chem. Rev. **120**, 4222 (2020).
[216] S. Hammes-Schiffer, J. Chem. Phys. **155**, 030901 (2021).
[217] H.J. Ache, Angew. Chem. Int. Ed. **11**, 179 (1972).
[218] D.C. Walker, *Muon and Muonium Chemistry* (Cambridge University Press, 1983).
[219] D.C. Walker, Acc. Chem. Res. **18**, 167 (1985).
[220] E. Roduner, *The Positive Muon as a Probe in Free Radical Chemistry: Potential and Limitations of the MSR Techniques* (Springer-Verlag, Berlin Heidelberg, 1988).
[221] C.M. Surko and F.A. Gianturco, editors, *New Directions in Antimatter Chemistry and Physics*, 2001st edition (Springer, Dordrecht ; Boston, 2001).
[222] C.J. Rhodes, J. Chem. Soc., Perkin Trans. 2 1379 (2002).
[223] Y.C. Jean, P.E. Mallon, and D.M. Schrader, *Principles and Applications of Positron and Positronium Chemistry* (World Scientific, 2003).
[224] E. Roduner, in *Isotope Effects In Chemistry and Biology*, edited by A. Kohen and H.-H. Limbach (CRC Press, 2005).
[225] N.J. Clayden, Phys. Scr. **88**, 068507 (2013).

[226] D. Horváth, in *Handbook of Nuclear Chemistry*, edited by A. Vértes, S. Nagy, Z. Klencsár, R.G. Lovas, and F. Rösch (Springer US, Boston, MA, 2011), pp. 1485–1513.
[227] W. von Oertzen, M. Freer, and Y. Kanada-En'yo, Phys. Rep. **432**, 43 (2006).
[228] M. Freer, Rep. Prog. Phys. **70**, 2149 (2007).
[229] A.K. Rai, J.N. Pandya, and P.C. Vinodkummar, Nucl. Phys. A **782**, 406 (2007).
[230] A.P. jr Mills, Riv. Del Nuovo Cim. **34**, 151 (2011).
[231] P. Bicudo, N. Cardoso, and M. Cardoso, Prog. Part. Nucl. Phys. **67**, 440 (2012).
[232] A.M. Frolov, Eur. Phys. J. D **72**, 222 (2018).
[233] M. Freer, H. Horiuchi, Y. Kanada-En'yo, D. Lee, and U.-G. Meißner, Rev. Mod. Phys. **90**, 035004 (2018).
[234] A. Ali, L. Maiani, and A.D. Polosa, *Multiquark Hadrons*, 1st edition (Cambridge University Press, 2019).
[235] R.F.G. Ruiz and A.R. Vernon, Eur. Phys. J. A **56**, 136 (2020).
[236] P. Nasertayoob, M. Goli, and S. Shahbazian, Int. J. Quantum Chem. **111**, 1970 (2011).
[237] M. Goli and S. Shahbazian, Int. J. Quantum Chem. **111**, 1982 (2011).
[238] F. Heidar Zadeh and S. Shahbazian, Int. J. Quantum Chem. **111**, 1999 (2011).
[239] M. Goli and S. Shahbazian, Theor. Chem. Acc. **129**, 235 (2011).
[240] M. Goli and S. Shahbazian, Theor. Chem. Acc. **131**, 1208 (2012).
[241] M. Goli and S. Shahbazian, Theor. Chem. Acc. **132**, 1362 (2013).
[242] M. Goli and S. Shahbazian, Theor. Chem. Acc. **132**, 1365 (2013).
[243] M. Goli and S. Shahbazian, Theor. Chem. Acc. **132**, 1410 (2013).
[244] M. Goli and S. Shahbazian, Phys. Chem. Chem. Phys. **16**, 6602 (2014).
[245] M. Goli and S. Shahbazian, Comput. Theor. Chem. **1053**, 96 (2015).
[246] M. Goli and S. Shahbazian, Phys. Chem. Chem. Phys. **17**, 245 (2015).
[247] M. Goli and S. Shahbazian, Phys. Chem. Chem. Phys. **17**, 7023 (2015).
[248] M. Goli and S. Shahbazian, Chem. Eur. J. **22**, 2525 (2016).
[249] M. Goli and S. Shahbazian, ChemPhysChem **17**, 3875 (2016).
[250] M. Gharabaghi and S. Shahbazian, J. Chem. Phys. **146**, 154106 (2017).
[251] M. Goli and S. Shahbazian, Phys. Chem. Chem. Phys. **20**, 16749 (2018).
[252] M. Goli and S. Shahbazian, ChemPhysChem **20**, 831 (2019).
[253] S. Shahbazian, Found. Chem. **15**, 287 (2013).
[254] P. Cassam-Chenaï and D. Jayatilaka, Theor. Chem. Acc. **105**, 213 (2001).
[255] R.F.W. Bader and P.F. Zou, Chem. Phys. Lett. **191**, 54 (1992).
[256] J. Charry, M.T. do N. Varella, and A. Reyes, Angew. Chem. Int. Ed. **57**, 8859 (2018).
[257] F. Moncada, L. Pedraza-González, J. Charry, M.T. do N. Varella, and A. Reyes, Chem. Sci. **11**, 44 (2019).
[258] S. Ito, D. Yoshida, Y. Kita, and M. Tachikawa, J. Chem. Phys. **153**, 224305 (2020).
[259] D. Bressanini, J. Chem. Phys. **154**, 224306 (2021).
[260] D. Bressanini, J. Chem. Phys. **155**, 054306 (2021).
[261] K. Ruedenberg, Rev. Mod. Phys. **34**, 326 (1962).
[262] W. Kutzelnigg, Angew. Chem. Int. Ed. **12**, 546 (1973).
[263] W. Kutzelnigg, Angew. Chem. Int. Ed. **23**, 272 (1984).
[264] A.M. Frolov and D.M. Wardlaw, Eur. Phys. J. D **63**, 339 (2011).
[265] T. Yamazaki and Y. Akaishi, Phys. Rev. C **76**, 045201 (2007).
[266] T. Yamazaki, Prog. Theor. Phys. Suppl. **170**, 138 (2007).
[267] J.M. Rost and D. Wintgen, Phys. Rev. Lett. **69**, 2499 (1992).

[268] N. Jiang and D.M. Schrader, Phys. Rev. Lett. **81**, 5113 (1998).
[269] G.G. Ryzhikh, J. Mitroy, and K. Varga, J. Phys. B: At. Mol. Opt. Phys. **31**, 3965 (1998).
[270] S.L. Saito, Nucl. Instrum. Methods Phys. Res. B **171**, 60 (2000).
[271] F. Rolim, J.P. Braga, and J.R. Mohallem, Chem. Phys. Lett. **332**, 139 (2000).
[272] J. Mitroy, M.W.J. Bromley, and G.G. Ryzhikh, J. Phys. B: At. Mol. Opt. Phys. **35**, R81 (2002).
[273] D. Bressanini and G. Morosi, J. Chem. Phys. **119**, 7037 (2003).
[274] L.J. Dunne and J.N. Murrell, Int. J. Quantum Chem. **96**, 512 (2004).
[275] F. Rolim, T. Moreira, and J.R. Mohallem, Braz. J. Phys. **34**, 1197 (2004).
[276] J. Mitroy, Phys. Rev. Lett. **94**, 033402 (2005).
[277] D.B. Cassidy and A.P. Mills Jr, Nature **449**, 195 (2007).
[278] D.B. Cassidy, S.H.M. Deng, and A.P. Mills, Phys. Rev. A **76**, 062511 (2007).
[279] J.-Y. Zhang and J. Mitroy, Phys. Rev. A **76**, 014501 (2007).
[280] C.M. Surko, Nature **449**, 153 (2007).
[281] D. Assafrão, H.R.J. Walters, and J.R. Mohallem, Nucl. Instrum. Methods Phys. Res. B **266**, 491 (2008).
[282] R. Heyrovska, Nat. Preced. 1 (2011).
[283] D.B. Cassidy, T.H. Hisakado, H.W.K. Tom, and A.P. Mills, Phys. Rev. Lett. **108**, 133402 (2012).
[284] D. Bressanini, Phys. Rev. A **97**, 012508 (2018).
[285] D. Bressanini, Phys. Rev. A **99**, 022510 (2019).
[286] M. Emami-Razavi and J.W. Darewych, Eur. Phys. J. D **75**, 188 (2021).
[287] W. Greiner, J.Y. Park, and W. Scheid, *Nuclear Molecules* (World Scientific Publishing Company, Singapore ; River Edge, N.J, 1995).
[288] A.H. Wuosmaa, R.R. Betts, M. Freer, and B.R. Fulton, Annu. Rev. Nucl. Part. Sci. **45**, 89 (1995).
[289] M. Freer and A.C. Merchant, J. Phys. G: Nucl. Part. Phys. **23**, 261 (1997).
[290] R.R. Betts and A.H. Wuosmaa, Rep. Prog. Phys. **60**, 819 (1997).
[291] B.R. Fulton, Contemp. Phys. **40**, 299 (1999).
[292] M. Freer, C. R. Phys. **4**, 475 (2003).
[293] H. Horiuchi, K. Ikeda, and K. Katō, Prog. Theor. Phys. Suppl. **192**, 1 (2012).
[294] M. Freer and H.O.U. Fynbo, Prog. Part. Nucl. Phys. **78**, 1 (2014).
[295] B. Zhou, Y. Funaki, H. Horiuchi, and A. Tohsaki, Front. Phys. **15**, 14401 (2019).
[296] M. Freer, Nature **487**, 309 (2012).
[297] J. Okołowicz, W. Nazarewicz, and M. Płoszajczak, Fortschr. Phys. **61**, 66 (2013).
[298] J.-P. Ebran, E. Khan, T. Nikšić, and D. Vretenar, Phys. Rev. C **87**, 044307 (2013).
[299] J.-P. Ebran, E. Khan, T. Nikšić, and D. Vretenar, J. Phys. G: Nucl. Part. Phys. **44**, 103001 (2017).
[300] R.B. Wiringa, S.C. Pieper, J. Carlson, and V.R. Pandharipande, Phys. Rev. C **62**, 014001 (2000).
[301] J.-P. Ebran, E. Khan, T. Nikšić, and D. Vretenar, Nature **487**, 341 (2012).
[302] R.J.N. Phillips, Rep. Prog. Phys. **22**, 562 (1959).
[303] M.J. Moravcsik, Rep. Prog. Phys. **35**, 587 (1972).
[304] R. Machleidt, Phys. Rep. **242**, 5 (1994).
[305] R. Machleidt and I. Slaus, J. Phys. G: Nucl. Part. Phys. **27**, R69 (2001).
[306] E. Epelbaum, H.-W. Hammer, and U.-G. Meißner, Rev. Mod. Phys. **81**, 1773 (2009).

[307] H. Joypazadeh and S. Shahbazian, Found. Chem. **16**, 63 (2014).
[308] S. Shahbazian, in *Applications of Topological Methods in Molecular Chemistry*, edited by R. Chauvin, C. Lepetit, B. Silvi, and E. Alikhani (Springer International Publishing, 2016), pp. 89–100.